\documentclass[pre,twocolumn,amsmath,amssymb]{revtex4}

\usepackage[T1]{fontenc}
\usepackage[utf8]{inputenc}
\usepackage{lmodern}
\usepackage[english]{babel}
\usepackage{xspace,cancel}

\usepackage{amsmath, amssymb, graphics, graphicx}
\usepackage{textcomp}
\usepackage{subfigure}
\usepackage{siunitx}
\usepackage{xcolor}
\usepackage[page,titletoc]{appendix}

\newcommand{\D}{{\mathrm d}}
\newcommand{\E}{\mathrm{e}}
\newcommand{\ii}{\mathrm{i}}

\newcommand{\LL}{\mathcal{L}}

\newcommand{\mP}{\mathcal{P}}

\newcommand{\average}[1]{\left<{#1}\right>}

\newcommand{\bx}{\mathbf{x}}

\newcommand{\bt}{\boldsymbol{\theta}}

\newcommand{\bm}{\mathbf{m}}

\newcommand{\p}[1]{\left({#1}\right)}
\newcommand{\pq}[1]{\left[{#1}\right]}

\DeclareSIUnit\stiffness{\pico\newton/\micro\meter}
\DeclareSIUnit\kbT{k_{\text{B}}T}

\definecolor{linkcolor}{rgb}{0,0,0.6}
\usepackage[pdftex,colorlinks=true, pdfstartview=FitV, linkcolor= linkcolor, citecolor= linkcolor, urlcolor= linkcolor, hyperindex=true,hyperfigures=true]{hyperref}

\begin{document}
\title{\bf Theoretical Description of Effective Heat Transfer between Two Viscously Coupled Beads}
\author{A. B\'{e}rut$^1$,  A. Imparato$^2$, A. Petrosyan$^1$,  S. Ciliberto $^1$}

\affiliation{$1 \ - $ Universit\'{e} de Lyon, CNRS, \\
Laboratoire de Physique, \'{E}cole Normale Sup\'{e}rieure de Lyon  (UMR5672),\\ 46 All\'{e}e d'Italie 69364 Lyon Cedex 07, France\\
\\
$2 \ - $ Department of Physics and Astronomy, University of Aarhus\\ Ny Munkegade, Building 1520, DK--8000 Aarhus C, Denmark}

\date{\today}

\begin{abstract}
We analytically study the role of non-conservative forces, namely viscous couplings, on the statistical properties of the energy flux between two Brownian particles kept at different temperatures. From the dynamical model describing the system, we identify an energy flow that satisfies a Fluctuation Theorem both in the stationary and in the transient state. In particular, for the specific case of a linear non-conservative interaction, we derive an exact Fluctuation Theorem that holds for any measurement time in the transient regime, and which involves the energy flux alone. Moreover, in this regime the system presents an interesting asymmetry between the hot and the cold particle. The theoretical predictions are in good agreement with the experimental results already presented in our previous article~\cite{Berut_PRL}, where we investigated the thermodynamic properties of two Brownian particles, trapped with optical tweezers, interacting through a dissipative hydrodynamic coupling.
\end{abstract}

\maketitle




\section{Introduction}

The heat transfer between two reservoirs kept at different temperatures, is certainly the simplest and probably the most fundamental out-of-equilibrium phenomenon that one can study. However, in small systems where the effects of thermal fluctuations on the mean heat flux cannot be neglected, this situation has been analyzed mainly in theoretical models~\cite{Deridda,Jarz2004,VandenBroeck,Visco,evans_temp,Villamania,Zamponi,Sarracino,Dotsenko, alb1,Fogedby12,Fogedby14}. Only a few very recent experiments have studied heat flux fluctuations~\cite{nostroPRL,ref:Ciliberto,Pekola_1,Berut_PRL}, because of the intrinsic difficulties of dealing with large temperature differences at small scales. 

In particular, in our recent paper~\cite{Berut_PRL} we have presented for the first time some measurements on the energy flux in a system with non conservative interactions. We have reported the statistical properties of the fluctuating energy transfer in a system composed by two trapped Brownian beads kept at different effective temperatures, and interacting only through hydrodynamic coupling.

Motivated by our previous experimental findings, in this article we derive the Fluctuation Theorems (FTs) characterizing the ``effective heat'' transfer between two Brownian particles interacting with non-conservative linear forces, both in the transient and stationary regime. This is an important problem in the study of the thermodynamics of small motors and Brownian ratchets, because fluctuating energy fluxes have been mostly investigated in systems with conservative couplings~\cite{Visco,evans_temp,Villamania,alb1,Fogedby12,Fogedby14}, and their statistical properties for dissipative interactions have been less discussed. For example they were addressed in ref.~\cite{Zamponi}, for a very specific dissipative coupling which is, however, difficult to implement in a real experiments. On the contrary, the set-up considered in~\cite{Berut_PRL} is rather general, as small devices immersed in a fluid often interact through hydrodynamic coupling (see for example~\cite{Coupling1,Coupling2,Coupling3,Coupling4}).

In the stationary regime, we find that energy fluxes satisfy a stationary exchange fluctuation theorem (xFT):
\begin{equation}
\ln \left( \frac{P(Q_{\tau})}{P(-Q_{\tau})} \right) \underset{\tau \to \infty}{=} \left( \frac{1}{k_{\text{B}} T_1}-\frac{1}{k_{\text{B}} T_2} \right) Q_{\tau}
\end{equation}
where $P(Q_{\tau})$ is the probability that an amount of (effective) heat $Q_{\tau}$ is exchanged during a time $\tau$ between the two systems at (effective) temperatures $T_1$ and $T_2$.
In the transient regime we find an asymmetry between the energy fluxes after the sudden application of the effective temperature gradient: while the hot-cold flux satisfies the transient xFT for any integration time, the energy flux between the cold and the hot particle obeys a FT only asymptotically for long times. In particular, for the specific case of the hydrodynamic linear coupling, we derive an exact xFT for the exchanged heat that holds at any time in the transient regime, and which generalizes to the non-conservative case the analogous FT discussed in~\cite{Imparato2014} for conservative interactions. We also show that some results for the conservative coupling case are recovered in the specific configuration where the two traps have the same stiffness. All these theoretical predictions show a good agreement with the experimental results reported in our previous article~\cite{Berut_PRL}.\bigskip

The article is organized as follow: we first present the model for the beads dynamics in section (\ref{sub:hydro_model}) and the definitions of the (effective) heat fluxes (\ref{sub:heat_definition}), and we show that the viscous coupling may produce non trivial effects in the energy balance (\ref{subsub:theory_mean_heat}). We then compute, in the framework of stochastic thermodynamics, the statistical properties of the heat flux both in the stationary (\ref{subsub:ss}) and transient state (\ref{subsub:tr}), and we show that the transient heat flux has non symmetric statistical properties.


\section{The hydrodynamic model}
\label{sub:hydro_model}

\begin{figure}[ht!]
\begin{center}
\includegraphics[width=0.5\textwidth]{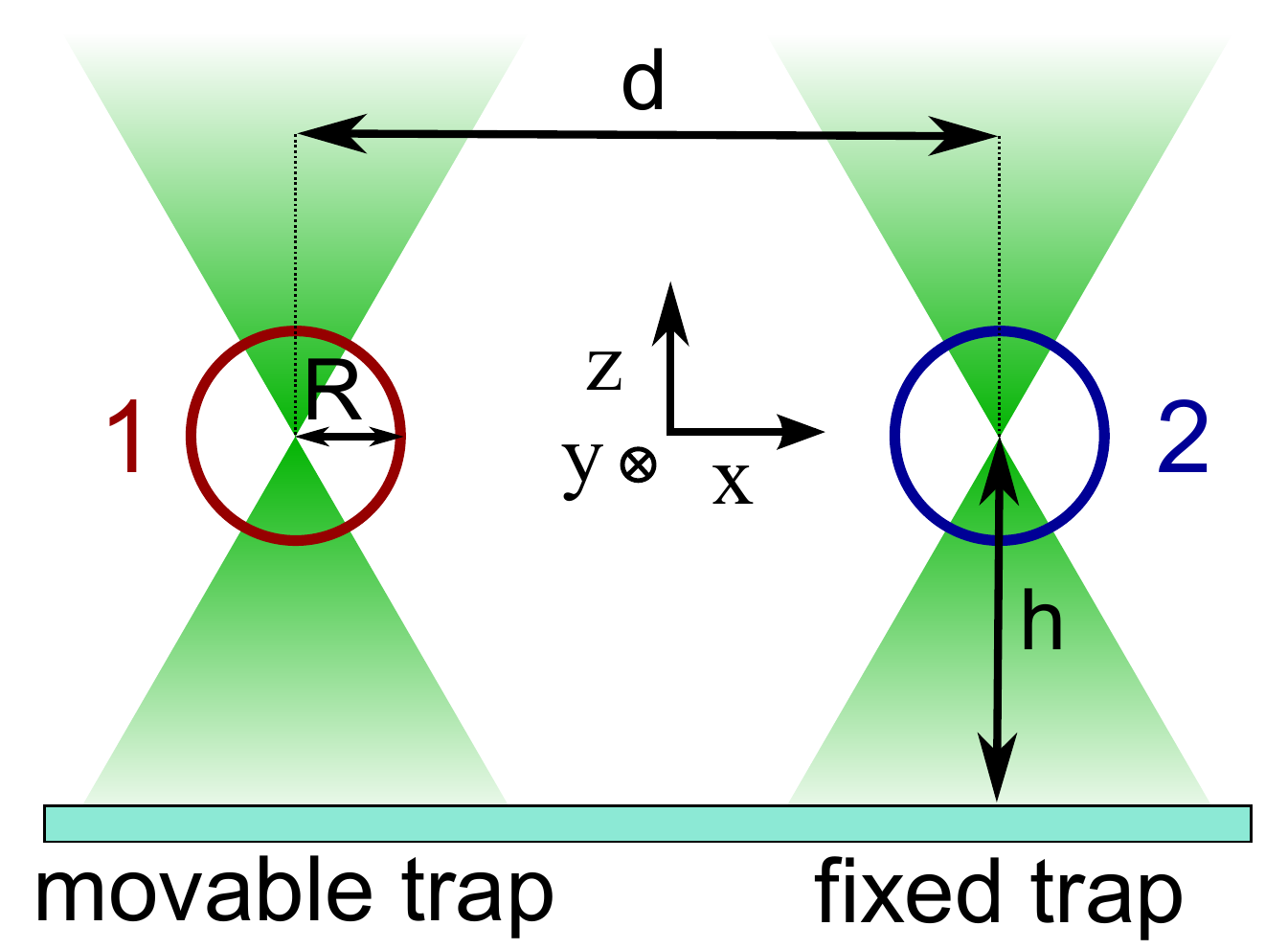}
\caption{Schematic representation of the two particles trapped by optical tweezers, separated by a distance $d$
along the $x$ axis.}
\label{fig:schematic_trap}
\end{center}
\end{figure}

Before introducing the stochastic equations describing the system dynamics, we briefly depict the experimental setup studied in ~\cite{Berut_PRL}, to show the connections between the present theoretical analysis and the experimental system.
To study a heat flux in presence of a dissipative coupling we used a system (schematically represented in figure~\ref{fig:schematic_trap}) composed of two spherical micro-particles, trapped in bi-distilled water by optical tweezers, and interacting only through an hydrodynamic coupling. One of the two particles is submitted to an external white noise, obtained by randomly displacing the position of its trap, which creates an ``effective temperature'' for the particle. The technical details about the experimental set-up can be found in~\cite{Berut_EPL,Berut_PRL}. 

Such a  system can be described by a classical hydrodynamic coupling model in low Reynolds-number flow. Following \cite{ref:Quake,ref:Bartlett,ref:Ou-Yang} the thermally excited motion of two identical particles of radius $R$ trapped at positions separated by a distance $d$ is described by two coupled Langevin equations:
\begin{equation}
\left( \begin{array}{c} \dot{x}_1 \\ \dot{x}_2 \end{array} \right) = 
\mathcal{H}
 \left( \begin{array}{c} F_1 \\ F_2 \end{array} \right)
\end{equation}
where $\mathcal{H}$ is the hydrodynamic coupling tensor, $x_i$ is the position of the particle $i$ relative to its trapping position and $F_i$ is the force acting on the particle $i$.

\noindent In the case where the displacements are small compared to the mean distance between the particles (i.e. $x \ll d$), the hydrodynamic coupling tensor reads:
\begin{equation}
\mathcal{H} = 
\begin{pmatrix} 1/\gamma & \epsilon/\gamma \\
 \epsilon/\gamma & 1/\gamma \end{pmatrix}
\end{equation}
where $\gamma$ is the Stokes friction coefficient ($\gamma = 6 \pi R \eta$ where $\eta$ is the viscosity of water) and $\epsilon$ is the coupling coefficient ($\epsilon = \frac{3R}{2d}$ if one takes the first order of the Oseen tensor \cite{ref:Doi}, $\epsilon = \frac{3R}{2d} - \left(\frac{R}{d}\right)^3$ if one takes the Rotne-Prager diffusion tensor \cite{ref:Herrera}).

\noindent It is important to notice that in the following theoretical treatment $\epsilon$ is considered as a constant because we consider only regimes where the fluctuations of particles positions are smaller than the mean distance between them.

At equilibrium the forces acting on the particles are:
\begin{equation}
F_i = -k_i \, x_i + f_i
\end{equation}
where $k_i$ is the stiffness of the trap $i$ and $f_i$ are the Brownian random forces which verify:
\begin{equation}
	\begin{array}{c}
		\langle f_i(t) \rangle = 0 \\
		\langle f_i(t) f_j(t^{\prime}) \rangle = 2k_{\mathrm{B}}T \, (\mathcal{H}^{-1})_{ij} \, \delta(t-t^{\prime})
	\end{array}
\end{equation}
where $k_{\mathrm{B}}$ is the Boltzmann constant and $T$ the temperature of the surrounding fluid.

To take account of a gradient of temperature between the two particles, we simply add an external random force $f^*$ on the first one. We make the assumption that this force is completely uncorrelated with the equilibrium Brownian random forces and that it is characterized by an additional temperature $\Delta T$ (the particle $1$ is then at a temperature $T^* = T + \Delta T$).
\begin{equation}
	\begin{array}{ccl}
		\langle f^*(t) \rangle &=& 0,\\ 
		\langle f^*(t) f_i(t^{\prime}) \rangle &=& 0,\\
		\langle f^*(t)f^*(t^{\prime}) \rangle &=& 2k_{\mathrm{B}} \Delta T \gamma \, \delta(t-t^{\prime}),
	\end{array}
	\label{eq:f_effective}
\end{equation}
\noindent It follows that the system of equations is:

\begin{eqnarray}
    \gamma \dot{x}_1 &=& -k_1 x_1 + \epsilon (-k_2 x_2 + f_2) + f_1 + f^*, \label{eqx1}\\
    \gamma \dot{x}_2 &=& -k_2 x_2 + \epsilon (-k_1 x_1 + f_1 + f^*) + f_2.\label{eqx2}
\end{eqnarray} 

In Eqs.\ref{eqx1} and \ref{eqx2}, the equilibrium Brownian noises, induced by the thermal bath verify:
\begin{eqnarray}
\langle f_i(t) \rangle &=& 0 \nonumber \\
\langle f_i(t) f_i(t^{\prime}) \rangle &=& \frac{2k_{\mathrm{B}} T \gamma}{1-\epsilon^2} \, \delta(t-t^{\prime})\nonumber \\
\langle f_1 (t) f_2(t^{\prime}) \rangle &=& - \epsilon \frac{2k_{\mathrm{B}} T \gamma}{1-\epsilon^2}  \, \delta(t-t^{\prime}) \nonumber
\end{eqnarray}
while the additional noise verifies equations \ref{eq:f_effective}.

\noindent It is important to notice that these relations are fully verified in the experimental set-up where $f^*$ is due to an external displacement of the first particle's trap. However, they might be modified if the additional noise on
the first particle were correlated with the equilibrium Brownian random forces.

Let us introduce the total normalised noises $ \xi_i$ acting on particle $i$, which read
\begin{eqnarray}
\xi_1 & = & \frac{1}{\gamma} ( \epsilon f_2 + f_1 + f^* ),\\
\xi_2 & = & \frac{1}{\gamma} ( \epsilon f_1 + \epsilon f^* + f_2 ).
\end{eqnarray}
We can thus rewrite eqs.~(\ref{eqx1})--(\ref{eqx2}) as 
\begin{equation}
\left\{
  \begin{array}{l}
    \dot{x}_1 = g_1(x_1,x_2)+\xi_1 \\
    \dot{x}_2 = g_2(x_1,x_2)+\xi_2 ,
  \end{array}
\right.
\label{lang:eq}
\end{equation}
with the normalised force reading
\begin{equation}
g_i(x_i,x_j)=-\frac 1 \gamma k_i x_i -\frac \epsilon \gamma k_j x_j,
\end{equation} 
and the normalised noises exhibiting the fluctuation-dissipation relations 
\begin{equation}
\average{\xi_i(t)\xi_j(t')}=2 \theta_{ij} \delta(t-t'),\label{fdr}
\end{equation} 
where we have introduced the diffusion matrix $\theta_{ij}$, whose elements read
\begin{eqnarray}
\theta_{11}&=&(T+\Delta T)/\gamma,\\
\theta_{12}&=&\epsilon (T+\Delta T)/\gamma,\\
\theta_{22}&=& (T+\epsilon^2 \Delta T)/\gamma.
\end{eqnarray} 

\section{The heat fluxes and their statistical properties}
\label{sub:heat_definition}
We now introduce the integrated heat $Q_i$ by analogy with the stochastic heat dissipated by a Brownian particle in a thermal bath~\cite{Sekimoto:Book}. $Q_i$ is defined as the work done by the stochastic forces on the particle $i$ in a time interval $\tau$:
\begin{eqnarray}
Q_i&=&\int_t^{t+\tau}(\gamma \dot x_i -\gamma \xi_i)\dot x_i\, \D t'\nonumber\\
&=& -\int_t^{t+\tau}(k_i  x_i +\epsilon k_j x_j)\dot x_i \, \D t',
\label{Qi:def}
\end{eqnarray} 
where the first equality holds in general, while the second one follows upon substitution of eq.~(\ref{lang:eq}).
Equation (\ref{Qi:def}) can be rewritten as 
\begin{equation}
Q_i=Q_{ii}+Q_{ij},
\label{sumQ}
\end{equation} 
where the heat $Q_i$ is split  into two contributions 
\begin{eqnarray}
Q_{ii}&=& -\int_t^{t+\tau}k_i  x_i \dot x_i \, \D t',  \label{Q12_1}\\
Q_{ij}&=& -\int_t^{t+\tau}\epsilon k_j x_j\dot x_i \, \D t',   \label{Q12}
\end{eqnarray} 
with $ i=\lbrace 1,2 \rbrace, \; j=\lbrace 2,1 \rbrace$.
While the first quantity is the work done on the particle $i$ by the quadratic trap, 
the latter quantity represents the work done on the particle $i$ by the non conservative force alone, due to the fluid-mediated particle-particle interaction.

Note that the integrals appearing in eqs.~(\ref{Q12_1})-(\ref{Q12}) are stochastic integrals which have to be interpreted according to the Stratonovich integration scheme.

\subsection{The mean values of the heat fluxes} 
\label{subsub:theory_mean_heat}

In this section we consider the total heat as  introduced in eq.~(\ref{sumQ}), calculate its average value, and  show that only the term involving the non--conservative force $Q_{ij}$  contributes to the average value $\average{Q_i}$.

Indeed from eq.~(\ref{Qi:def}) we have
\begin{eqnarray}
\dot Q_i&=&-(k_ix_i+\epsilon k_j x_j) \dot x_i=-(k_ix_i+\epsilon k_j x_j)(g_i+\xi_i)\nonumber \\
&=&F_{Q_i}+\xi_{Q_i},\label{dotQi}
\end{eqnarray} 
which is a Langevin-like equation for the stochastic variable $Q_i(t)$.
The deterministic force acting on $Q_i(t)$ reads 
\begin{equation}
F_{Q_i}=-(k_ix_i+\epsilon k_j x_j) g_i=\frac 1 \gamma (k_ix_i+\epsilon k_j x_j)^2,
\label{fq:det}
\end{equation} 
while the stochastic force reads
\begin{equation}
\xi_{Q_i}=-(k_ix_i+\epsilon k_j x_j) \xi_i.
\end{equation} 

In appendix~\ref{appendix_FP}, starting from the Langevin equations (\ref{lang:eq}) and (\ref{dotQi}), we derive the Fokker-Plank equation for the probability distribution $P(x_1,x_2,t)$ and  for the joint probability distribution $P(x_1,x_2,Q_i,t)$, which allow us to calculate the heat rates
\begin{eqnarray}
q_1&=&\partial_t \average{Q_1}=\frac{k_{\mathrm{B}} \Delta T k_2^2 \epsilon ^2 \left(\epsilon ^2-1\right)}{\gamma  (k_1+k_2)}\label{q1:def},\\
q_2&=&\partial_t \average{Q_2}=-\frac{k_{\mathrm{B}} \Delta T k_1 k_2 \epsilon ^2 \left(\epsilon ^2-1\right)}{\gamma  (k_1+k_2)}\label{q2:def}
\end{eqnarray} 
with $q_1=q_2=0$ if $\Delta T=0$, and $q_1+q_2=0$ if $k_1=k_2$.

\begin{figure}[!ht]
\begin{center}
\subfigure[]{ \includegraphics[width=0.46\linewidth]{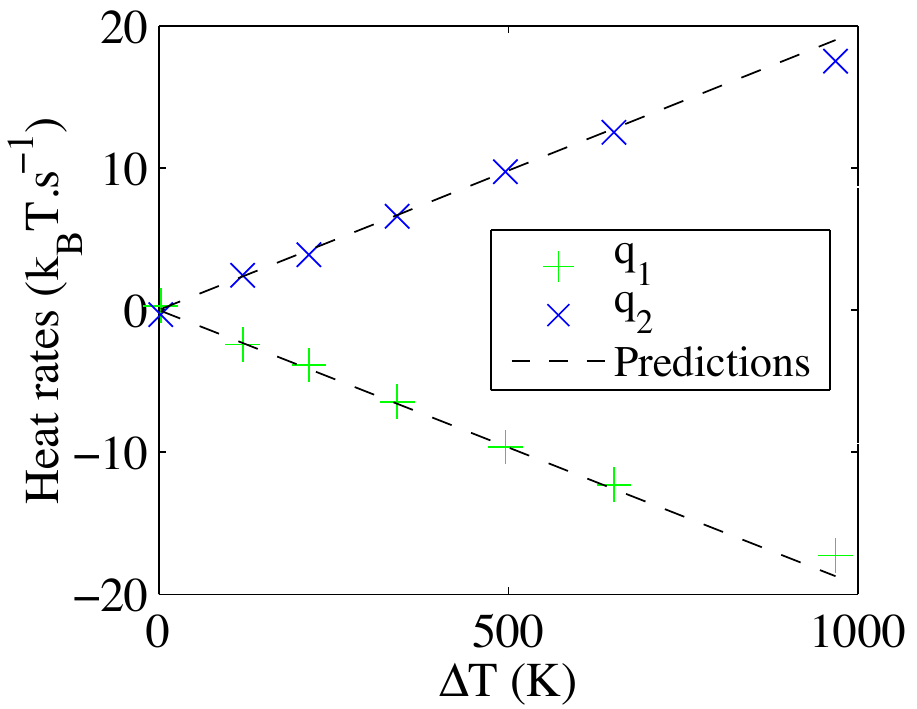} }
\subfigure[]{ \includegraphics[width=0.46\linewidth]{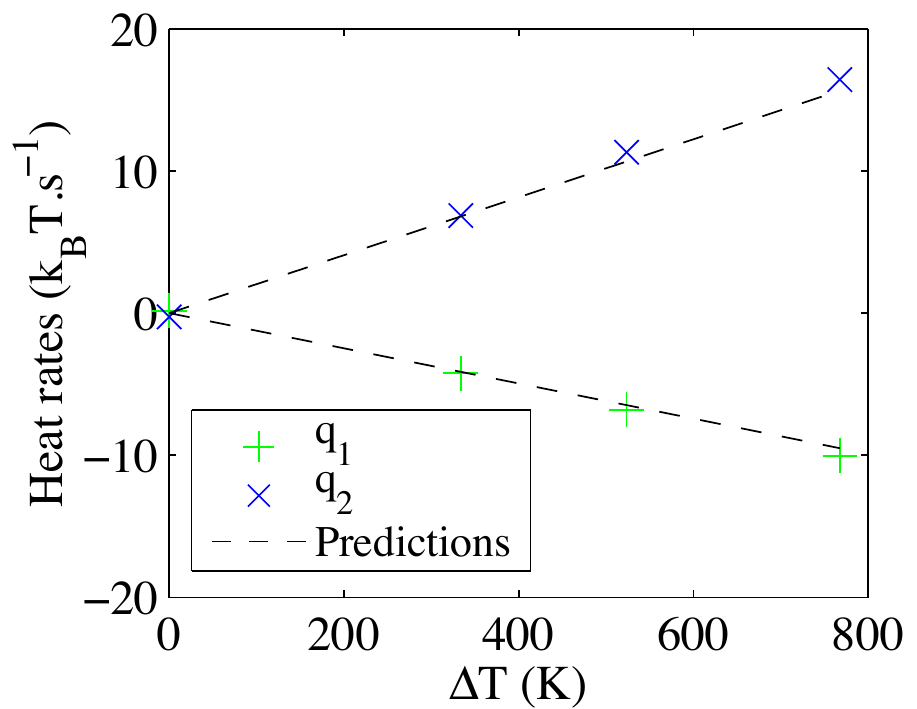} }
\caption{Experimental heat rates $q_1$ and $q_2$.  a) In the case $k_1 = k_2$  ($\approx \SI{3.35}{\stiffness}$) b) In the case $k_1\ne k_2$ ($k_1 \approx \SI{4.20}{\stiffness}$ and $k_2 \approx \SI{2.55}{\stiffness}$). The dashed lines  are the theoretical predictions obtained by substituting the corresponding experimental values in equations~(\ref{q1:def}) and~(\ref{q2:def}).}
\label{fig:Mean_heat_prediction}
\end{center}
\end{figure}

Using data from the experiments presented in~\cite{Berut_PRL}, we can evaluate the experimental values of the heat rate $q_i$ (see figure~\ref{fig:Mean_heat_prediction}). They show a good agreement with the values predicted by eqs.~(\ref{q1:def}) and~(\ref{q2:def}).

We now turn our attention to the stochastic variables $Q_{ij}$, and consider the Langevin equation expressing their time evolution, which reads
\begin{equation}
\dot Q_{ij}=-\epsilon k_j x_j \dot x_i=-\epsilon k_j x_j(g_i+\xi_i)=F_{Q_{ij}}+\xi_{Q_{ij}},\label{dotQij} 
\end{equation} 
where the deterministic force acting on $Q_{ij}(t)$ is 
\begin{equation}
F_{Q_{ij}}=-\epsilon k_j x_j g_i=\frac 1 \gamma  \pq{\epsilon k_ix_i k_jx_j+(\epsilon k_j x_j)^2},
\end{equation} 
while the stochastic force reads
\begin{equation}
\xi_{Q_{ij}}=-\epsilon k_j x_j \xi_i.
\end{equation} 
Following the same steps as above (see appendix~\ref{appendix_FP} for the details) one finds the rates 
\begin{eqnarray}
q_{12}&=&\partial_t \average{Q_{12}}=\frac{k_{\mathrm{B}} \Delta T k_2^2 \epsilon ^2 \left(\epsilon ^2-1\right)}{\gamma  (k_1+k_2)} \label{q12:rate}\\
q_{21}&=&\partial_t \average{Q_{21}}=-\frac{k_{\mathrm{B}} \Delta T k_1 k_2 \epsilon ^2 \left(\epsilon ^2-1\right)}{\gamma  (k_1+k_2)}
\label{q21:rate}
\end{eqnarray} 
which are identical to eqs.~(\ref{q1:def})-(\ref{q2:def}).
By recalling the definitions of $Q_i, \, Q_{ii}$ and $Q_{ij}$, eqs.(\ref{sumQ})--(\ref{Q12}), we conclude that the heat rates $q_1$ and $q_2$ are the work done per time unit by the non-conservative forces alone.
This makes sense if one considers the following physical argument: the heat flows because of the term $x_2$ on the right hand side of the equation for $\dot x_1$, and because of the term $x_1$ on the right hand side of the equation for $\dot x_2$ in the Langevin equations (\ref{lang:eq}). Then $Q_{ij}$ can be seen as the work done by the interaction force on each particle. This would be the case even if the coupling forces were conservative. 

We also notice that $q_1=-q_2$ only when $k_1=k_2$ because in this case the system is perfectly symmetric, and the equations~(\ref{lang:eq}) become equivalent to those of a system with a conservative coupling:
\begin{equation}
\left\{
  \begin{array}{l}
    \gamma \dot{x}_1 = -\partial_{x_1}U  + \xi_1, \\
    \gamma \dot{x}_2 = -\partial_{x_2}U + \xi_2,
  \end{array}
\right.
\end{equation}\newline
where $U(x_1,x_2)=k \left(x_1^2 + 2 \epsilon x_1 x_2 + x_2^2 \right)/2$. 
In this case, we have the energy conservation $Q_1(t)+Q_2(t)=U(x_1(t),x_2(t))-U(x_1(0),x_2(0))$ for any time $t$. On the other hand $q_1\ne-q_2$ when  $k_1\ne k_2$ because in the general case the forces are not conservative, and thus we cannot invoke the conservation of energy.
Thus the fact that the heat dissipated by particle 1 is not the opposite of the heat dissipated by particle 2 is only due to the dissipative nature of the coupling, and it is not induced by any specificity of the random forcing $f^*$. In the experimental system, if $k_1=k_2$,  the dissipative nature of the coupling cannot be observed and the energy fluxes $Q_i$ due to the ``effective temperature difference'' behave as conventional heat fluxes in conservative systems.

\section{Steady state and transient Fluctuation Theorems}
\label{sub:FT_theory}

In the present section we present the main results concerning the FTs, while the detailed proofs are discussed in appendix~\ref{appendix_FT}.

Starting from the Langevin equations for the microscopic variables (\ref{lang:eq}), one can calculate the probability of observing a single trajectory in the phase space. Given an initial  $\bx=(x_1,x_2)$ and a final state $\bx'=(x'_1,x'_2)$, the transition probability $P_F(\bx \rightarrow \bx'| \bx, t)$ of a trajectory connecting the points  $\bx$ and  $\bx'$ in a finite time interval $[t,t,+\tau]$ and  the transition probability $P_B(\bx' \rightarrow \bx| \bx', t+\tau)$ of the corresponding backward trajectory obey the detailed FT
\begin{eqnarray}
\frac{P_F(\bx \rightarrow \bx'| \bx, t)}{P_B(\bx' \rightarrow \bx| \bx', t+\tau)} &=& \E^{ \frac{1}{k_{\mathrm{B}}} \p{\frac{Q_{11}}{T+\Delta T} +\frac{ Q_{22}}{T}-\frac{\Delta T }{T(T+\Delta T)}Q_{12}}}
\label{pfpb:main}
\end{eqnarray} 
where the quantities $Q_{ij}$ are given by eqs.~(\ref{Q12_1})-(\ref{Q12})  and depend on the specific trajectory. It is worth to note that this equation holds for any $\tau$.

\subsection{The steady state}
\label{subsub:ss}
Starting from eq.~(\ref{pfpb:main}) we can prove  the steady state FTs for the quantities $Q_i$ and $Q_{ij}$, see appendix \ref{appendix_FT_ss} for details.
Specifically we prove that in the steady state,  $Q_{12}$ and $Q_{21}$ verify an FT  in the long time limit : 
\begin{equation}
 \ln \left( \frac{P(Q_{12})}{P(-Q_{12})} \right) \underset{\tau \to \infty}{\simeq}  \left( \frac{1}{k_{\text{B}} T_1}-\frac{1}{k_{\text{B}} T_2} \right) Q_{12}
\label{eq:FT_Q12_t}
\end{equation}
and 
\begin{equation}
 \ln \left( \frac{P(Q_{21})}{P(-Q_{21})} \right)  \underset{\tau \to \infty}{\simeq} \frac{k_2}{k_1} \left( \frac{1}{k_{\text{B}} T_2}-\frac{1}{k_{\text{B}} T_1} \right) Q_{21}.
\label{eq:FT_Q21_t}
\end{equation}
Here we have rewritten  eqs.~\ref{eq:FT_Q12},~\ref{eq:FT_Q21} of appendix~\ref{appendix_FT_ss} to have  the explicit dependence on $T_1$ and $T_2$.

These xFT are analogous to the one presented in~\cite{Jarz2004} for the heat exchanged between two heat bath put in contact during a time $\tau$. It is interesting to notice that because of the dissipative coupling the FT for $Q_{21}$ has a prefactor $k_2/k_1$ which disappears only in the symmetric case when $k_2=k_1$.

\subsection{The transient regime}
\label{subsub:tr}

Following the protocol discussed in \cite{Fogedby14, Imparato2014}, we now assume that we prepare our system such that  at $t\to -\infty$ the temperature difference is vanishing $\Delta T=0$, and then at $t=0$ we suddenly turn on the temperature difference $\Delta T$, with $T_1=T+\Delta T$, and start measuring the heat currents for $t\ge 0$.
We assume to prepare the system in the same way along the backward trajectories: at $t'=\tau-t\to -\infty $ we take  $\Delta T=0$, and then  at $t'=0$ we ``turn on" the temperature difference $\Delta T$ and start measuring the heat currents along the backward trajectories.
Since the particle interaction do not change in time, the PDF for the  initial  position of the forward trajectory is equal to the PDF for the  initial  position of the backward trajectory.
Starting from eq.~(\ref{pfpb:main}), in appendix \ref{appendix_FT_trans} we prove that, for the  specific protocol where the system is prepared with $\Delta T=0$ at the beginning of both the forward and the backward trajectories, the following detailed FT holds for any $\tau$
\begin{eqnarray}
\frac{P(x_1,x_2,t=0) P_F(\bx \rightarrow \bx'| \bx, 0)}{P(x_1',x_2',t'=0) P_B(\bx' \rightarrow \bx| \bx', \tau)}
&=&\E^{-\frac{\Delta T}{k_{\mathrm{B}} T( T+\Delta T)} Q_1}. \label{FT:fin:main}
\end{eqnarray} 
This FT generalizes to the non-conservative case the analogous FT discussed in~\cite{Imparato2014} for conservative interactions.
As discussed in \cite{Imparato2014} such a FT involves only the measurable heat currents $\dot Q_i$ and no boundary term depending on the final and the initial state $\bx'$ and $\bx$ appears in its expressions, at variance with the FT discussed in \cite{Seifert05a}. However, differently from \cite{Imparato2014}, in the present study we consider non-conservative forces, and thus the condition of local detailed balance does not apply here.
Had we chosen to change suddenly the temperature $T_2$ at $t=0$, with $T_2=T+\Delta T$, the heat flux $Q_2$ would appear on the rhs of eq.~(\ref{FT:fin:main}) instead of $Q_1$.

As a consequence the PDF of $Q_1$ obeys the integrated FT
\begin{equation}
 \ln \left( \frac{P(Q_1,\tau)}{P(-Q_1,\tau)} \right) = \left( \frac{1}{k_{\text{B}} T_1}-\frac{1}{k_{\text{B}} T_2} \right) Q_{1}
, \label{FT:fin:int_t}
\end{equation} 
which  holds for any $ \tau>0$, while the heat $Q_2$ verifies an FT
\begin{equation}
 \ln \left( \frac{P(Q_2,\tau)}{P(-Q_2,\tau)} \right)\underset{\tau \to \infty}{\simeq} \frac{k_2}{k_1} \left( \frac{1}{k_{\text{B}} T_2}-\frac{1}{k_{\text{B}} T_1} \right) Q_2 \label{FT:fin2:int_t}.
\end{equation} 
which holds only in the limit of large $\tau$, see appendix \ref{appendix_FT_trans} for the details.\bigskip

As already mentioned in introduction, the theoretical predictions for the statistical properties of the heat have already been compared with the experimental results both in the stationary and transient regimes~\cite{Berut_PRL}.

\section{Discussion and conclusions} \label{sec:conclusion}

This article presents several theoretical results on the energy exchanged between two Brownian particles coupled by viscous interactions, and kept at different temperatures. 

Starting from the coupled Langevin equations, we have defined a heat flux as the energy exchanged between a particle at equilibrium with the fluid, and a particle submitted to an uncorrelated additional source of noise.
For this energy flow, we have theoretically demonstrated the following behaviors:
\begin{itemize}
\item[a)] The mean fluxes are linear functions of the effective temperature difference. 
\item[b)] The exchanged fluxes $Q_{ij}$ satisfy the exchange fluctuation theorem (xFT) in the stationary state.
\item[c)] The total flux for the hot particle $Q_1$ satisfies the transient xFT for any time after the sudden application of the temperature gradient, whereas the total flux for the cold particle $Q_2$ satisfies it only asymptotically (i.e. for long integration times).
\end{itemize}
In particular, the last property had been previously predicted only for systems with a conservative coupling~\cite{Fogedby14} and we have proved it also in the case of dissipative linear coupling. For all the theoretical predictions, the experimental results show a very good agreement. Note also that the integrated FTs in points b) and c) follow from more general detailed FTs that hold at the level of single trajectories.
Our results show that there are strong analogies between the statistical properties of a dissipatively coupled system with those of a conservatively coupled one (see also ref.~\cite{Berut_JSTAT_2016} for a detailed comparison), and are particularly relevant in all of the cases where an external random forcing is applied to a system which is coupled to another one.

\section{Acknowledgments}
We acknowledge very useful discussion with K. Sekimoto. This work has been partially supported by the ERC project OUTEFLUCOP.   AI is supported by the  Danish Council for Independent Research, the Villum Fonden, and the COST Action MP1209 ``Thermodynamics in the Quantum Regime''. 

\newpage

\begin{appendices}

\numberwithin{equation}{section}

In these appendices we take  $k_{\mathrm{B}}=1$ to simplify the notation. 

\section{The Fokker-Planck equations for the probability distribution of the stochastic variables} \label{appendix_FP}

Starting from the Langevin equation (\ref{lang:eq}) for the the stochastic variables $x_i$,
we introduce the Fokker-Planck (FP) equation for the probability distribution function (PDF) $P(x_1,x_2,t)$, which reads \cite{Zwanzig2001}
\begin{eqnarray}
&&\partial_tP(x_1,x_2,t)=\LL_0P(x_1,x_2,t)\label{LL0} \\
&=&-\partial_{x_i}\pq{g_i(x_2,x_2)P(x_1,x_2,t)}+\partial_{x_i}\partial_{x_j}\theta_{ij} P(x_1,x_2,t).\nonumber
\end{eqnarray} 
The stationary steady state PDF, obeying $\partial_t P(x_1,x_2,t)=0$, reads
\begin{equation}
P_{ss}(x_1,x_2,\Delta T)=\frac{\sqrt{ac -b^2}}{\pi} \exp\pq{-\p{ax_1^2+2 b x_1 x_2 +c x_2^2}},
\label{pdfss}
\end{equation} 
where the dependence on the temperature gradient $\Delta T$ has been made explicit, as it will be useful in the following, and 
with 
\begin{eqnarray}
a&=&\frac{k_1 (k_1+k_2)\pq{(k_1+k_2)T +\epsilon^2 k_2 \Delta T}}{\Theta},\nonumber\\
b&=&-\frac{\epsilon k_1 k_2 (k_1+k_2)\Delta T}{\Theta},\nonumber\\
c&=&\frac{k_2 (k_1+k_2)\pq{(k_1+k_2)(T+\Delta T) -\epsilon^2 k_2 \Delta T}}{\Theta},\nonumber\\
\Theta&=&2  \pq{(T^2+T \Delta T)(k_1+k_2)^2-\epsilon^2(\epsilon^2-1)k_2^2 \Delta T^2}.\nonumber
\end{eqnarray} 
From eq.~(\ref{pdfss}) we also notice that the steady state PDF for  a vanishing temperature gradient reads 
\begin{equation}
P_{ss}(x_1,x_2,\Delta T=0)\propto \exp\pq{- \p{\frac{k_1}{2 T} x_1^2 +  \frac{k_2}{2  T} x_2^2 }}\label{pss0}.
\end{equation} 
as expected because the coupling is dissipative. 

We now consider the Langevin equation for $Q_i$ as given by eq.~(\ref{dotQi}) in the main text. The stochastic forces acting on the variables $x_i(t)$ (eq.~(\ref{lang:eq})) and on $Q_i(t)$ (eq.~(\ref{dotQi})) obey the 
fluctuation-dissipation relations
\begin{eqnarray}
\average{\xi_i(t) \xi_{Q_i}(t')}&=&-(k_ix_i+\epsilon k_j x_j)2 \theta_{ii} \delta(t-t'),\nonumber \\
\average{\xi_j(t) \xi_{Q_i}(t')}&=&-(k_ix_i+\epsilon k_j x_j)2 \theta_{ij} \delta(t-t'),\nonumber \\
\average{\xi_{Q_i}(t) \xi_{Q_i}(t')}&=&(k_ix_i+\epsilon k_j x_j)^2 2 \theta_{ii} \delta(t-t').\nonumber
\end{eqnarray} 
The Fokker-Planck (FP) equation for the joint probability distribution $\mP(x_1,x_2,Q_i,t)$ reads \cite{Imparato07,Imparato08,Fogedby12,Fogedby14}
\begin{widetext}
\begin{eqnarray}
\partial_t \mP&=&\LL_0\mP-\partial_{Q_i} \p{F_{Q_i} \mP}-\theta_{ii} \pq{(k_ix_i+\epsilon k_j x_j)\partial_{x_i}\partial_{Q_i} \mP+ \partial_{x_i}\partial_{Q_i} (k_ix_i+\epsilon k_j x_j) \mP}\nonumber \\
&& -\theta_{ij} \pq{(k_ix_i+\epsilon k_j x_j)\partial_{x_j}\partial_{Q_i} \mP+ \partial_{x_j}\partial_{Q_i} (k_ix_i+\epsilon k_j x_j) \mP} + \gamma F_{Q_i} \theta_{ii}\partial^2_{Q_i} \mP,
\label{FP:eq}
\end{eqnarray} 
\end{widetext}
where $\LL_0$ is the FP operator for the variables $x_1$ and $x_2$ alone appearing in eq.~(\ref{LL0}).
We now introduce the generating function 
\begin{equation}
\psi(x_1,x_2,\lambda,t)=\int \D Q_i \mP(x_1,x_2,Q_i,t) \E^{\lambda Q_i},
\label{gen:eq}
\end{equation} 
and from  (\ref{FP:eq}) we obtain the FP equation for  $\psi(x_1,x_2,\lambda,t)$
\begin{widetext}
\begin{eqnarray}
\partial_t \psi&=&\LL_0\psi+ \lambda F_{Q_i} \psi+ \lambda \theta_{ii} \pq{(k_ix_i+\epsilon k_j x_j)\partial_{x_i} \psi+ \partial_{x_i} (k_ix_i+\epsilon k_j x_j) \psi}\nonumber \\
&& +\lambda \theta_{ij} \pq{(k_ix_i+\epsilon k_j x_j)\partial_{x_j} \psi+ \partial_{x_j} (k_ix_i+\epsilon k_j x_j) \psi} + \gamma F_{Q_i} \theta_{ii}\lambda^2 \psi,
\label{FP:eq1}
\end{eqnarray}
\end{widetext}
From eq.~(\ref{gen:eq}) we notice that 
\begin{eqnarray}
\partial_t \average{Q_i}=\partial_t \int \D x_1 \D x_2\, \left. \partial_\lambda \psi(x_1,x_2,\lambda,t)\right|_{\lambda=0},
\end{eqnarray} 
and thus from eq.~(\ref{FP:eq1}) we obtain
\begin{equation}
\partial_t \average{Q_i}=\partial_t \int \D \bx\, \left. \partial_\lambda \psi(\bx,\lambda,t)\right|_{\lambda=0}=\average{F_{Q_i}}-\theta_{ii} k_i-\theta _{ij} \epsilon k_j.
\label{derQ}
\end{equation} 

By noticing that $\average{F_{Q_i}}=\average{(k_ix_i+\epsilon k_j x_j)^2}/\gamma$ (see eq.~(\ref{fq:det})), and using eq.~(\ref{pdfss}) to calculate the correlations $\average{x_{i}^2}$ and $\average{x_i x_j}$, one obtains the heat rates $q_1$ and $q_2$, as given in eqs.~(\ref{q1:def})-(\ref{q2:def}) in the main text.

The same procedure is used to obtain the rates for the quantities $\average{Q_{ij}}$ as given by eqs.~(\ref{q12:rate})-(\ref{q21:rate}) in the main text.

\section{Proof of the Fluctuation Theorem} \label{appendix_FT}

In order to proof the Fluctuation Theorem for the heat exchanged between two reservoirs in the case of dissipative coupling , we need  to calculate the probability of a single trajectory in the phase space. 
We first introduce 
the distribution of the noises $\xi_i$ appearing in eq.~(\ref{lang:eq}). Those are gaussian correlated noises with fluctuation-dissipation relations  given by eq.~(\ref{fdr}), and thus their probability distribution reads 
\begin{equation}
\phi(\xi_1,\xi_2)= \frac{\Delta t}{\pi} |\bm|^{1/2} \exp\pq{-\Delta t\p{m_{11} \xi_1 +2 m_{12} \xi_1 \xi_2+m_{22} \xi_2^2}}
\end{equation} 
where $\bm$ is the symmetric correlation matrix $\bm=\bt^{-1}/4$, with elements
\begin{eqnarray}
m_{11}&=&\frac{\gamma  \left(\Delta T \epsilon ^2+T\right)}{4 T \left(1-\epsilon ^2\right) (T+\Delta T)},\\
m_{22}&=&\frac{\gamma}{4 T(1-\epsilon^2)},\\
m_{12}&=&-\frac{\gamma}{4 T(1-\epsilon^2)},
\end{eqnarray} 
and  $\Delta t$ is a small time increment.

We now calculate the transition probability from a state $\bx=(x_1,x_2)$ to a new state $\bx'=(x'_1,x'_2)$ in a time interval $\Delta t$. Let $\Delta x_i=x'_i-x_i$, we have 
\begin{widetext}
\begin{eqnarray}
P_F(\bx \rightarrow \bx'| \bx, t)&=&\int\D \xi_1\D \xi_2\,  \delta(\Delta x_1-\Delta t(g_1(x_1,x_2)+\xi_1)) \delta(\Delta x_2-\Delta t(g_2(x_1,x_2)+\xi_2)) \phi(\xi_1,\xi_2)\nonumber \\
&=&\int \D q_1\D q_2\E^{\ii q_i \Delta x_i}\int \D \xi_1\D \xi_2\, \E^{\Delta t \pq{q_i (g_i(x_i,x_j)+\xi_i) + m_{ij} \xi_i \xi_j}}\nonumber\\
&=&\frac{4 \pi}{\Delta t}  |\bm|^{1/2} \, \E^{ -\frac{\Delta t } { m_{22} } \left[|\bm| (  \dot x_1-g_1(x_1,x_2))^2+(m_{22} ( \dot x_2-g_2(x1,x2))+m_{12} (\dot  x_1-g_1(x_1,x_2)))^2\right]},
\end{eqnarray} 
where the sum over repeated indexes is understood.
Similarly, we found for the reverse (backward) transition
\begin{eqnarray} 
P_B(\bx' \rightarrow \bx| \bx', t+\Delta t)&=&\int\D \xi_1\D \xi_2\,  \delta(\Delta x_1+\Delta t(g_1(x_1',x_2')+\xi_1)) \delta(\Delta x_2+\Delta t(g_2(x_1',x_2')+\xi_2)) \phi(\xi_1,\xi_2)\nonumber \\
&=&\frac{4 \pi}{\Delta t}  |\bm|^{1/2} \, \E^{ -\frac{\Delta t } { m_{22} } \left[|\bm| (  \dot x_1+g_1(x'_1,x'_2))^2+(m_{22} ( \dot x_2+g_2(x'_1,x'_2))+m_{12} (\dot  x_1+g_1(x'_1,x'_2)))^2\right]},
\end{eqnarray} 
\end{widetext}

By taking the ratio between the forward and backward probability, we find 
\begin{eqnarray}
&&\frac{P_F(\bx \rightarrow \bx'| \bx, t)}{P_B(\bx' \rightarrow \bx| \bx', t+\Delta t)}=\nonumber\\
&&=\E^{-\frac{\Delta t (k_1 T x_1 \dot x_1 +k_2 x_2  ((T+\Delta T) \dot x_2-\Delta T \epsilon\dot x_1  ))}{T (T+\Delta T)}}\nonumber \\
&&= \E^{\Delta t \p{\frac{\dot Q_{11}}{T+\Delta T} +\frac{\dot Q_{22}}{T}-\frac{\Delta T }{T(T+\Delta T)}\dot Q_{12}}}
\label{pfpb}
\end{eqnarray} 
where $Q_{ij}$ is given by eq.~(\ref{Q12}).

By iterating the above procedure, and considering a trajectory from $\bx$ to $\bx'$ in a finite time interval $[t,t,+\tau]$, from eq.~(\ref{pfpb}), we find
\begin{eqnarray}
&&\frac{P_F(\bx \rightarrow \bx'| \bx, t)}{P_B(\bx' \rightarrow \bx| \bx', t+\tau)}=\\
&=&\E^{- \int_{t}^{t+\tau}\D t' \p{\frac{ k_1 x_1 \dot x_1}{T+\Delta T} +\frac{ k_2 x_2  \dot x_2}{T}-\frac{\Delta T}{T(T+\Delta T)} \epsilon k_2 x_2\dot x_1 }} \label{pfpb:fin}\\
&=& \E^{- \frac{k_1({x_1'}^2-x_1^2)}{2 (T+\Delta T)}  -  \frac{k_2}{2 T} ({x_2'}^2-x_2^2)-  Q_{12} \frac{\Delta T} {T (T+\Delta T)}} \label{pfpb_tau}\\
&=&  \E^{- \frac{k_1({x_1'}^2-x_1^2)}{2 (T+\Delta T)}  -  \frac{k_2}{2 T} ({x_2'}^2-x_2^2)+\frac{k_2\Delta T(Q_{21}/k_1+\epsilon  (x_1'x_2'-x_1x_2))} {T (T+\Delta T)} },\label{pfpb:fin1}
\end{eqnarray} 
where we have used $Q_{12}=-k_2(Q_{21}/k_1+\epsilon (x_1'x_2'-x_1x_2))$. Eq.~(\ref{pfpb:fin}) corresponds to eq.~(\ref{pfpb:main}) in the main text.
The last set of equations hold at the level of single trajectories, and for any time interval $\tau$.

\section{Steady state FT} \label{appendix_FT_ss}
Here we derive the  FTs introduced in section \ref{subsub:ss}.

Equation (\ref{pfpb:fin}) gives the long time FT for the quantity $Q_{12}$. Indeed by noticing that the quantity $Q_{12}$ scales linearly with the time on average, as discussed above (see eq.~(\ref{q12:rate}) in the main text), while the differences ${x_i'}^2-x_i^2$ are time independent, we can write the following approximate equality  
\begin{eqnarray}
\frac{P_F(\bx \rightarrow \bx'| \bx, t)}{P_B(\bx' \rightarrow \bx| \bx', t+\tau)}\underset{\tau \to \infty}{\simeq}\E^{-  Q_{12} \frac{\Delta T} {T (T+\Delta T)} }
\label{ft:approx} 
\end{eqnarray}
which is the FT for $Q_{12}$ for large $\tau$. This corresponds to neglecting the boundary terms in in eq.~(\ref{pfpb_tau}), i.e. energy difference stored in the potentials $k_i x_i^2/2$. By noticing that if the quantity $Q_{12}$ is associated to a given forward trajectory, then  $-Q_{12}$ is associated to the corresponding backward  trajectory, and by integrating the lhs of eq.~(\ref{ft:approx}) over all those trajectories with a fixed value of $Q_{12}$, and neglecting the contribution of any initial distribution  of the variable $\bx$ for the forward trajectories and of $\bx'$ for the backward trajectories, we obtain the long-time integrated FT for $Q_{12}$ which reads
\begin{equation}
\frac{P(Q_{12},\tau)}{P(-Q_{12},\tau)}\underset{\tau \to \infty}{\simeq}\E^{-  Q_{12} \frac{\Delta T} {T (T+\Delta T)} }.
\label{eq:FT_Q12}
\end{equation} 

Similarly eq.~(\ref{pfpb:fin1}) gives the FT for the quantity $Q_{21}$, indeed by neglecting the boundary terms, one obtains
\begin{eqnarray}
\frac{P_F(\bx \rightarrow \bx'| \bx, t)}{P_B(\bx' \rightarrow \bx| \bx', t+\tau)}\underset{\tau \to \infty}{\simeq}\E^{  \frac{k_2 }{k_1} Q_{21}\frac{\Delta T} {T (T+\Delta T)} },
\label{ft:approx1} 
\end{eqnarray}
while the  long-time integrated FT for $Q_{21}$  reads
\begin{equation}
\frac{P(Q_{21},\tau)}{P(-Q_{21},\tau)}\underset{\tau \to \infty}{\simeq}\E^{ \frac{k_2 }{k_1} Q_{21} \frac{\Delta T} {T (T+\Delta T)} }.
\label{eq:FT_Q21}
\end{equation} 

\section{The transient FT} \label{appendix_FT_trans}

We prepare our system such that at $t\to -\infty$ the temperature difference is vanishing $\Delta T=0$, and then at $t=0$ we ``turn on" the temperature difference $\Delta T$, and start measuring the heat currents for $t\ge 0$.
Thus the initial PDF for our system reads 
\begin{equation}
P(x_1,x_2,t=0)=P_{ss}(x_1,x_2,\Delta T=0),
\end{equation} 
where $P_{ss}(x_1,x_2,\Delta T=0)$ is given by eq.~(\ref{pss0}).
We prepare the system in the same way along the backward trajectories: at $t'=\tau-t\to -\infty $ we take  $\Delta T=0$, and then  at $t'=0$ we ``turn on" the temperature difference $\Delta T$ and start measuring the heat currents along the backward trajectories.
Thus we have 
\begin{eqnarray}
&&\frac{P(x_1,x_2,t=0)}{P(x'_1,x'_2,t'=0)}=\frac{P_{ss}(x_1,x_2,\Delta T=0)}{P_{ss}(x_1',x_2',\Delta T=0)}=\nonumber\\
&& =\E^{\frac{k_1}{2 T} ({x_1'}^2-x_1^2) +  \frac{k_2}{2 T} ({x_2'}^2-x_2^2) }.
\end{eqnarray} 
Thus, combining the last equation with eq.~(\ref{pfpb:fin}), we obtain the following expression
\begin{eqnarray}
&&\frac{P(x_1,x_2,t=0) P_F(\bx \rightarrow \bx'| \bx, 0)}{P(x_1',x_2',t'=0) P_B(\bx' \rightarrow \bx| \bx', \tau)}=\nonumber\\
&&=\exp\pq{\frac{\Delta T}{T(T+\Delta T)} \int_0^{\tau} \D t'\,  k_1 x_1 \dot x_1+\epsilon k_2 x_2\dot x_1} \nonumber \\
&&=\exp\pq{-\frac{\Delta T}{T( T+\Delta T)} Q_1}, \label{FT:fin}
\end{eqnarray} 
which is a FT that holds for any $ \tau>0$ and for the specific protocol where the system is prepared with $\Delta T=0$ at the beginning of both the forward and the backward trajectories. The last equation corresponds to eq.~(\ref{FT:fin:main}) in the main text. 

We now derive the integrated FTs introduced in section \ref{subsub:tr}.
Upon integration over all the microscopic trajectories with a fixed value of $Q_1$, eq.~(\ref{FT:fin}) gives the integrated FT
\begin{equation}
\frac{P(Q_1,\tau)}{P(-Q1,\tau)}=\exp\pq{-\frac{\Delta T}{T( T+\Delta T)} Q_1}, \label{FT:fin:int}
\end{equation} 
which also holds for any $ \tau>0$.

We consider now the explicit expression of $Q_1$ and  $Q_2$, and using eq.~(\ref{Qi:def}), we express  these quantities quantities as
\begin{eqnarray}
Q_1&=&\frac {k_2}{2}({x_1'}^2-x_1^2)+\epsilon k_2 \int_{t}^{t+\tau}\D t' x_2\dot x_1,\nonumber\\
Q_2&=&\frac {k_2}{2}({x_2'}^2-x_2^2) +\epsilon k_1 \int_{t}^{t+\tau}\D t' x_1\dot x_2\nonumber \\
&=&\frac {k_2}{2}({x_2'}^2-x_2^2)\nonumber\\
&&+\epsilon k_1 \pq{(x'_1 x'_2-x_1x_2)-\int_{t}^{t+\tau}\D t' x_2\dot x_1},\nonumber
\end{eqnarray} 
We can thus recast the expression of $Q_1$ in terms of $Q_2$, and find 
\begin{equation}
Q_1=-\frac{k_2}{k_1}Q_2+ \frac{k_1}{2 }({x_1'}^2-x_1^2)+ \frac{k_2^2}{2k_1}({x_2'}^2-x_2^2)+ \epsilon k_2 (x'_1 x'_2-x_1x_2).\nonumber
\end{equation} 
By substituting the last equation into eq.~(\ref{FT:fin}) and neglecting the boundary terms in the long time limit,  we find   
\begin{eqnarray}
\frac{P(x_1,x_2,t=0) P_F(\bx \rightarrow \bx'| \bx, 0)}{P(x_1',x_2',t'=0) P_B(\bx' \rightarrow \bx| \bx', \tau)}
\underset{\tau \to \infty}{\simeq}\E^{\frac{k_2 \Delta T}{k_1 T( T+\Delta T)} Q_2 },\nonumber
\end{eqnarray} 
and upon integration over  all the microscopic trajectories with a fixed value of $Q_2$ we find the integrated FT 
\begin{equation}
\frac{P(Q_2,\tau)}{P(-Q_2,\tau)}\underset{\tau \to \infty}{\simeq}\exp\pq{\frac{k_2 \Delta T}{k_1 T( T+\Delta T)} Q_2} \label{FT:fin2:int}.
\end{equation} 

\end{appendices}

\bibliographystyle{iopart-num}
\bibliography{biblio}
\end{document}